# *Pattern overlapping decomposition by Cumulative Local Cross-Correlation*

*Simon Kogan*
*Genome Diversity Center, Institute of Evolution, University of Haifa, Mount Carmel, Haifa 31905, Israel, Email: skogan@research.haifa.ac.il*

## *Abstract*

### Background

Nucleotide sequences contain multiple codes responsible for organism's functioning and structure. They can be investigated by various signal processing methods. These techniques are well suited for indication of frequently encountered sequence motifs (i.e., repeats). However, if there are two or more codes containing the same motif, the local nucleotide distribution (i.e., profile), resulting from sequence alignment by the motif position, will represent overlapping of the code patterns.

### Results

The novel algorithm for decomposition of pattern overlapping is proposed. It is capable to work with dispersed repeats as well. The algorithm is based on cross-correlation procedure applied locally in a cumulative fashion. Its sensitivity was tested on human genomic sequences.

### Conclusions

*Cumulative Local Cross-Correlation* was successfully used to decompose overlapping of nucleotide patterns in human genomic sequences. Being very general technique (as general as cross-correlation), it can be easily adopted in other signal processing applications and naturally extended for multidimensional cases.

Software implementation of the algorithm is available on request from the authors.

## *Background*

Nucleotide sequences, a store of biological inheritance information, contain multiple codes (messages) responsible for organism's functioning and structure. Nucleic acid-to-protein triplet code was discovered first [1]. Among others, known today with some degree of detailing, are: promoter sequences, DNA shape code, chromatin code, gene splicing code, the translation framing code, etc. [2-review, 3].

A nucleic sequence can be seen as a signal and investigated by various signal processing methods: Fourier [4] and wavelet [5, 6] transforms, cross-correlation analysis [7, 8], etc. These techniques are well suited for indication of common (frequently encountered) sequence motifs that are always suspected to be a part of some code associated with known or unknown cell functioning. A dispersed motif (frequent combination of oligonucleotides following each other at some specific distance with an arbitrary sequence in between) can be captured as well.

Calculation of local oligonucleotide distribution in the vicinity of the motif is quite a trivial task. One just has to align sequences by the motif position and sum number of the particular oligonucleotide in each position in the window enclosing the motif. If there is only one code containing this motif, the distribution (it can be nucleotide, dinucleotide or any other oligonucleotide distribution according to the nature of the code and the need of the research) is actually a pattern (profile) representing the code. It can be used directly for finding of repeats in new sequences or converted to a consensus sequence. However, if there are several different codes containing the motif, the distribution will represent an overlapping of all their patterns. To reconstruct one of the patterns or all of them, one has to decompose the distribution picture. To the best of our knowledge, none of the existing techniques provides a way to do this without usage of *a priori* knowledge about the pattern(s), similar to the classic blind deconvolution problem of signal and image processing (see [9-11] reviews).

The *Cumulative Local Cross-Correlation* algorithm (referred further as *CLCC*) developed in this work permits decomposing of pattern overlapping in a straightforward manner without assuming anything about the nature of the investigated code(s).

Section **2** of the paper provides the formal representation of *CLCC* illustrated by its application to a synthetic data. Section **3** shows the use of the technique on human genomic sequence. The final section discusses the applicability of the algorithm in various areas and its possible development and extension.

## *The Cumulative Local Cross-Correlation algorithm*

### Sequence motif definition

To describe dispersed sequence motifs, it is important to formalize the distance notion. In this paper, distances are measured in nucleotide units. If the distance between nucleotides is equal to $l$, there are $l-1$ other nucleotides between them. The motif comprised from 'C', 'T' and 'G' nucleotides in ACTAG sequence is represented as C-1-T-2-G. The distance between oligonucleotides is measured as the distance between their first nucleotides (for example, AC-3-AG represents the motif comprised from AC and AG dinucleotides in the above sequence).

### Synthetic data

In order to illustrate the technique, a random nucleotide sequence was constructed. Two nucleotide patterns: A-15-T-10-A and T-10-A-15-A-10-T were randomly added to the sequence. There is the same motif T-10-A belonging to both of the patterns. The cross-correlation procedure applied to this sequence will obviously reveal the motif. However, we expect that local nucleotide distributions in the vicinity of the motif will represent overlapping of the two patterns. We will show how the *Cumulative Local Cross-Correlation* technique can distinguish between them.

### Nucleotide cross-correlation and local distribution

Let's calculate nucleotide cross-correlation in this sequence. There are $4^2 = 16$ such functions (in this paper, the cross-correlation definition includes auto-correlation definition as well). To implement the calculation, one can represent the sequence by 4 discrete signals: the first one for 'A' nucleotide positions ('1' – where it is present, '0' – where it is absent); the second one for 'C' nucleotide positions, etc. The nucleotide cross-correlation formula is as follows:

$CC_{n1n2}(l) = \sum_{i=1}^{i=N-l} s_{n1}(i) * s_{n2}(i+l)$, where: $s_n(\tau)$ - a discrete signal; *'n1'* (and *'n2'*) – one of 'A', 'C', 'G' or 'T'; *'N'* – length of the signal (equal to length of the sequence); *'i'* – position in the signal; *'l'* – cross-correlation lag. Figure 1 shows a plot of the cross-correlation between 'T' and 'A' nucleotides. As we expected, the highest peak is found in lag 10 (representing the common motif for both of the patterns). Other 15 cross-correlations are not presented here.

Given the initial motif (T-10-A in our case), calculation of local oligonucleotide distribution by the motif position sequence alignment (see Introduction) is a trivial procedure. However, we provide here the formal definition of it to use in the *CLCC* algorithm's explanation.

The local (in the window) nucleotide distribution formula is as follows:

$d_n(k) = \sum_{i=wr+1}^{N-wr-(motifLen-1)} hit(i) * s_n(i+k)$, where: *'n'*, $s_n(\tau)$, *'i' and 'N'* – as above;

$hit(i) = \begin{cases} 1 - there_is_an_initial_motif_in_position_(i) \\ 0 - otherwise \end{cases}$ (motif position is defined, by convention, by the position of its first nucleotide); *motifLen* – length of the initial motif (equal to 11 for T-10-A); *'wr'* – window radius (Thus, window size = 2**wr*+*motifLen*.); *'k'* – position in the window ($-wr \le k \le (motifLen-1)+wr$ with $k=0$ corresponding to the first nucleotide of the initial motif).

Figure 2 shows 'A' and 'T' distributions in the vicinity of T-10-A initial motif found in the synthetic sequence. 'T' and 'A' are always present in position (0) and (10) respectively (just by the nature of the distribution's assembly). Other peaks A(-15), A(25) and T(35) belong to two added patterns. Given the distributions only, without *a-priory* knowledge or theory about the nature of the signal, it is impossible to distinguish between the patterns, because phase, representing spatial relation information, is lost (similar to a deconvolution problem). However, the *CLCC* algorithm reveals (by collecting additional information during the distribution assembly) the relationship between the peaks and, therefore, permits recovering of the pattern(s).

## The *CLCC* definition

For each position of initial motif ($hit(i)=1$), calculate cross-correlation between the signals in each pair of positions inside a window enclosing the motif. The result of this operation is a matrix (local cross-correlation instance) with elements given as follows:

$LCC_i^{n1n2}(k_1,k_2) = s_{n1}(i+k_1) * s_{n2}(i+k_2)$, where: $s_n(\tau)$, *'n1'*, *'n2'* and *'i'* – as above; $k_1$ and $k_2$ – matrix row and column indices respectively. Their range ($-wr \le k_1, k_2 \le (motifLen-1)+wr$) is the same as local distribution range (It is defined so for clarity sake. The actual matrix element indices, that have to be positive, are calculated as follows: $\langle k_1, k_2 \rangle + wr + 1$). The matrix width and height are, thus, equal to local distribution window size.

Calculate *Cumulative Local Cross-Correlation* matrix as the sum of local cross-correlation instances:

$$CLCC^{n1n2}(k_1,k_2) = \sum_{i=wr+1}^{N-wr-(motifLen-1)} hit(i) * LCC_i^{n1n2}(k_1,k_2).$$

The number of *CLCC* matrices is equal to the number of cross-correlation functions (16 in nucleotide case). Figure 3 shows AT-*CLCC* matrix for our synthetic example (the other *CLCC* matrices are not presented here). Each element of *CLCC* matrix characterizes correlation between pair of positions in original local distribution.

Now, one just needs to check a related *CLCC* element in order to know whether two specific peaks in the original distribution belong to the same pattern or not. If both of them do belong to one pattern, there will be a significant peak (relatively to background level) in this element. The peak height estimations in this chapter are the qualitative ones. The formal statistical significance calculation example is given in the next chapter.

Obviously, the high peak gives larger correlation value than the low one with every position in the window regardless of pattern relation. It creates horizontal, vertical and diagonal ridges and valleys in *CLCC* matrix. To eliminate the correlation estimation bias, the matrix is normalized by subtraction from all diagonals and rows their medians. Vertical ridges (valleys) do not impede the further analysis performed along the matrix columns.

Let's, inspect the peaks from Figure 2. The element related to the correlation between A(-15) and A(25) peaks is located in (-15) column and (25) row of AA-*CLCC* matrix. The plot of column (-15) of AA-*CLCC* (see Figure 4) shows no peak in position (25). Similar check for A(-15) and T(35) peaks in AT-*CLCC* reveals no peak as well. However, checking correlation between A(25) and T(35), we get a large peak significantly higher than the background noise level (see Figure 4).

Absence of correlation between A(-15) peak and other two peaks means that it belongs (together with the initial motif T-10-A) to the separate pattern consisting of three peaks: A(-15), T(0) and A(10) (it is shifted A-15-T-10-A). A(25) and T(35) peaks correlate with each other, and therefore, constitute (together with the initial motif) the second separate pattern: T(0), A(10), A(25), and T(35).

Thus, using the *CLCC* algorithm, we were able to reconstruct both of the patterns originally added to the synthetic sequence.

## The CLCC application to human genomic sequences

The following example illustrates the application of *CLCC* to one contig (7MB) of human chromosome **1**. Development of this algorithm took place during the investigation of a nucleosome positioning nucleic pattern based on dinucleotides [12]. Sequence, in this case, is represented by 16 discrete signals for dinucleotide positions ('1' – where it is present, '0' – where it is absent). The *CLCC* was adjusted to support it. The definitions are the same as above with the following exception: *'n'* (*'n1'*, *'n2'*) – one of AA, AC, …, TT. There are 16 local distributions and 256 cross-correlation functions (and *CLCC* matrices as well) in dinucleotide case.

To evaluate statistical significance of peak sizes, we generated artificial contigs from the original one by shuffling dinucleotide positions (with preservation of overall dinucleotide compositions). The sequences were processed in the same way as the original contig.

Cross-correlation analysis of the contig reveals a peak in the lag 12 in AA-TT cross-correlation (see Figure 5). The peak is highly statistically significant (p-value < 0.001). Thus, we calculated local dinucleotide distributions and *CLCC* matrices in the vicinity of AA-12-TT initial motif. Two of the distributions are shown in Figure 6.

To inspect the possibility of pattern overlapping, we checked (using *CLCC* technique) the correlation between two largest peaks in CG distribution in positions (-9) and (21) and a peak in AA distribution in position (40) (see Figure 7). The plot of CGAA-*CLCC* column (-9) indicates no peak in row (40) that means existence of at least two separate patterns. The CGAA-*CLCC* column (21) plot reveals a high (p-value < 0.0001) peak in row (40) that points to strong relation between CG(21) and AA(40) peaks.

Considering correlation between peaks of local distribution, one can construct a new motif as an arbitrary combination of mutually correlated peaks. However, it is bound to include the initial motif that compulsory correlates to all peaks. Obviously, the stronger the correlation the better (more prominent) is the motif. The presence of several separate groups of mutually correlated peaks indicates the existence of the same number of different patterns.

The motif that we constructed is as follows: AA-12-TT-9-CG-19-AA (thus, the distance from the first AA to CG is 21 and from the first AA to the second AA is 40). Dinucleotide local distributions in its vicinity were calculated as above (Figure 8 illustrates two of them). To prove that this distribution represents a non-overlapped pattern, one should check the mutual correlation of distribution's peaks by a new *CLCC* analysis. However, the consensus sequence, obtained from the distribution, was found to be very similar to the well-known Alu repeat widely present in human genome [13]. That made unnecessary the further analysis.

## Consensus sequence reconstruction

Obviously, only non-overlapped local oligonucleotide distributions are suitable for conversion to a repeat's consensus sequence. Otherwise, the outcome is meaningless. Priory to the conversion, oligonucleotide distributions have to be transformed to nucleotide ones. The conversion itself is straightforward: one just chooses the highest value among four nucleotide distributions in each position.

Continuing with human genome contig evaluation, we converted the dinucleotide distributions obtained with AA-12-TT-9-CG-19-AA motif to the consensus nucleotide sequence. Two points are worth mentioning here:

- In the conversion process we are interested in distribution peaks rather than the average distribution value (for example, AA dinucleotide is much more frequent than CG dinucleotide that causes the differences in distribution mean level (see Figure 6)). Therefore, the above distributions were normalized by dinucleotide occurrence before the conversion.

- The conversion robustness along the distribution was estimated by statistical significance of occurrence of the highest value nucleotide in each position. P-value = 0.05 was set as the significance level.

The alignment of the obtained sequence with Alu consensus [14] reveals that the sequence has longer 'A' run in the end (see Table 1). No full Alu sequence from Repbase [15] express such features. We attribute the result to the sensitivity of *CLCC* algorithm.

In order to extract the second pattern from the overlapped dinucleotide distributions in the vicinity of AA-12-TT initial motif, we could find distribution peak(s) correlated to CG(-9) peak and reconstruct a related pattern using the above techniques. However, taking into account the symmetry of CG self-complimentary dinucleotide distribution (see Figure6), it became clear that the distributions were dominated by overlapping of two patterns: Alu and complimentary Alu (both of them contain AA-12-TT motif).

### The algorithm's computational complexity

The *CLCC* matrix set can be built in polynomial time. The formal expression is: $O(n, w^2)$. It is linear to sequence length (n), and quadric to window size (w).

The memory consumption can be expressed as follows: $S(w^2, m)$; where $m$ – number of *CLCC* matrices that depends on size of a language and word length. With DNA's 4 letter language, it is $4^{2*oligoLength}$ (16 and 256 in nucleotide and dinucleotide cases respectively).

## Conclusions

Cross-correlation and consequently *Cumulative Local Cross-Correlation* are suitable when there is a chance to build a consensus sequence or profile (as with nucleosome positioning pattern). In the case of variable in length signal (like promoter), it may help if the signal is clusterizable into number of specific length signals.

With well conserved signals (like Alu), the technique permits actual building of consensus sequence or profile. In the case of very weak signals (like nucleosome positioning pattern), the technique serves as a tool for analysis and evaluation.

Theoretically, by checking all pairs of peaks in local distribution with *CLCC* technique and collecting groups of mutually correlated peaks, all patterns can be separated from each other (even if there are more than two patterns). However, the task becomes quite tedious (but not impossible) when several complex patterns create a lot of peaks in overlapped distribution picture. Actually, we designed an automatic procedure for finding and thresholding of *CLCC* peaks. Like any heuristics of this kind, it performs better when signal to noise ratio is high. Therefore, masking of strong signals (frequent and well conserved repeats) improves sensitivity to the weak ones.

The *Cumulative Local Cross-Correlation* is a very general technique (as general as cross-correlation itself). It was developed for the purpose of study of DNA sequences that influences the explanations and examples in a certain way. However, this relatively simple and mathematically well definable algorithm is readily deployed for any one-dimensional signal and can be easily extended for multi-dimensional signals as well.

## Acknowledgments

The work was partially supported by Caesarea Edmond Benjamin de Rothschild Foundation Institute for Interdisciplinary Applications of Computer Science (C.R.I.).

## References

1. Nirenberg M, Leder P, Bernfield M, Brimacombe R, Trupin J, Rottman F, O'Neal C: **RNA code words and protein synthesis, vii. On the general nature of the RNA code**. *Biochemistry* 1965, **53**.
2. Trifonov EN: **The multiple codes of nucleotide sequences**. *Bull Math Biol* 1989, **51**(4):417-432.
3. Trifonov EN: **Interfering contexts of regulatory sequence elements**. *Comput Appl Biosci* 1996, **12**(5):423-429.
4. Peng CK, Buldyrev SV, Goldberger AL, Havlin S, Sciortino F, Simons M, Stanley HE: **Long-range correlations in nucleotide sequences**. *Nature* 1992, **356**(6365):168-170.
5. Arneodo A, Bacry E, Graves PV, Muzy JF: **Characterizing long-range correlations in DNA sequences from wavelet analysis**. *Physical Review Letters* 1995, **74**(16):3293-3296.
6. Lio P: **Wavelets in bioinformatics and computational biology: state of art and perspectives**. *Bioinformatics* 2003, **19**(1):2-9.
7. Trifonov EN, Sussman JL: **The pitch of chromatin DNA is reflected in its nucleotide sequence**. *Proc Natl Acad Sci U S A* 1980, **77**(7):3816-3820.
8. Herzel H, Trifonov EN, Weiss O, Grosse I: **Interpreting correlations in biosequences**. *Physica A* 1998, **249**(1-4):449-459.
9. Jain AK: **Fundamentals of Digital Image Processing**, Paperback edn: Prentice Hall; 1 edition (September 23, 1988); 1988.
10. Stainvas I, Intrator N: **Blurred face recognition via a hybrid network architecture**. *Pattern Recognition, 2000 Proceedings* 2000, **2**:805-808.
11. Gonzalez RC, Woods RE: **Digital Image Processing**, Hardcover edn: Prentice Hall; 2nd edition (January 15, 2002) l; 2002.
12. Kato M, Onishi Y, Wada-Kiyama Y, Abe T, Ikemura T, Kogan S, Bolshoy A, Trifonov EN, Kiyama R: **Dinucleosome DNA of human K562 cells: experimental and computational characterizations**. *J Mol Biol* 2003, **332**(1):111-125.
13. Hwu HR, Roberts JW, Davidson EH, Britten RJ: **Insertion and/or deletion of many repeated DNA sequences in human and higher ape evolution**. *Proc Natl Acad Sci U S A* 1986, **83**(11):3875-3879.
14. Jurka J, Smith T: **A fundamental division in the Alu family of repeated sequences**. *Proc Natl Acad Sci U S A* 1988, **85**(13):4775-4778.
15. Jurka J: **Repbase update: a database and an electronic journal of repetitive elements**. *Trends Genet* 2000, **16**(9):418-420.

## *Figure and table legends*

**Figure 1** Cross-correlation between ‘T’ and ‘A’ nucleotides in the synthetic sequence. X-axis – correlation lag; y-axis – correlation count. The highest peak (marked by arrow) in lag 10 relates to the common motif T-10-A in both of the added patterns: A-15-T-10-A and T-10-A-15-A-10-T. The lower peak (marked by arrow) in lag 25 relates to the motif T-25-A in the second pattern.

**Figure 2** Distributions of ‘A’ (top) and ‘T’ (bottom) nucleotides in the vicinity of T-10-A motif found in the synthetic sequence. The highest peaks: in position (10) in ‘A’ distribution and position (0) in ‘T’ distribution relate to the motif. Other peaks (marked by arrows) relate to two added patterns.

**Figure 3** Cumulative local cross-correlation between ‘A’ and ‘T’ nucleotides in the vicinity of T-10-A motif found in the synthetic sequence. The highest ridge represents correlation of A(10) to the rest of the window. The high peak (marked by arrow) in the junction of A(25) column and T(35) row reveals the belonging of these positions to the same pattern.

**Figure 4** T-10-A motif’s *CLCC* matrix columns. On all graphs: x-axis – matrix row index; y-axis – value of normalized *CLCC* matrix elements. Top graph: A(-15) column of AA-*CLCC*. Peak in row (10) relates to A(10) of the motif. Middle graph: A(-15) column of AT-*CLCC*. Peak in row (0) relates to T(0) of the motif. Bottom graph: A(25) column of AT-*CLCC*. Peak in row (0) relates to T(0) of the motif. Peak in row (35) (marked by arrow) reveals belonging of A(25) and T(35) to the same pattern.

**Figure 5** Cross-correlation between AA and TT dinucleotides in one contig (7MB) of human chromosome **1**. X-axis – correlation lag; y-axis – correlation count. The highest peak (marked by arrow) is in lag 12.

**Figure 6** Distribution of AA (top) and CG (bottom) dinucleotides in the vicinity of AA-12-TT motif found in human genome contig. The highest peak in position (0) in ‘AA’ distribution relates to the motif. Three peaks (marked by arrows): AA(40), CG(-9) and CG(21) were analyzed by *CLCC* technique.

**Figure 7** AA-12-TT motif‘s CGAA-*CLCC* matrix columns. On all graphs: x-axis – matrix row index; y-axis – value of normalized *CLCC* matrix elements. Top graph: CG(-9) column indicates no peak in row (40). Bottom graph: CG(21) column has a high peak (marked by arrow) in row (40) that reveals the belonging of CG(21) and AA(40) to the same pattern. The highest peak in row (0) in both of the graphs relates to AA(0) of the motif.

**Figure 8** Distribution of ‘AA’ (top) and ‘CG’ (bottom) dinucleotides in the vicinity of AA-12-TT-9-CG-19-AA motif found in human genome contig. The highest peaks: in positions (0) and (40) in ‘AA’ distribution and position (21) in ‘CG’ distribution relate to the motif.

**Table 1** Alignment between Alu consensus and the consensus sequence obtained from human genome contig by *CLCC* application. ‘x’ marks places where conversion robustness was lower than 0.05 significance level. Adenine run in the end of the sequence is longer than the one in Alu consensus.

##

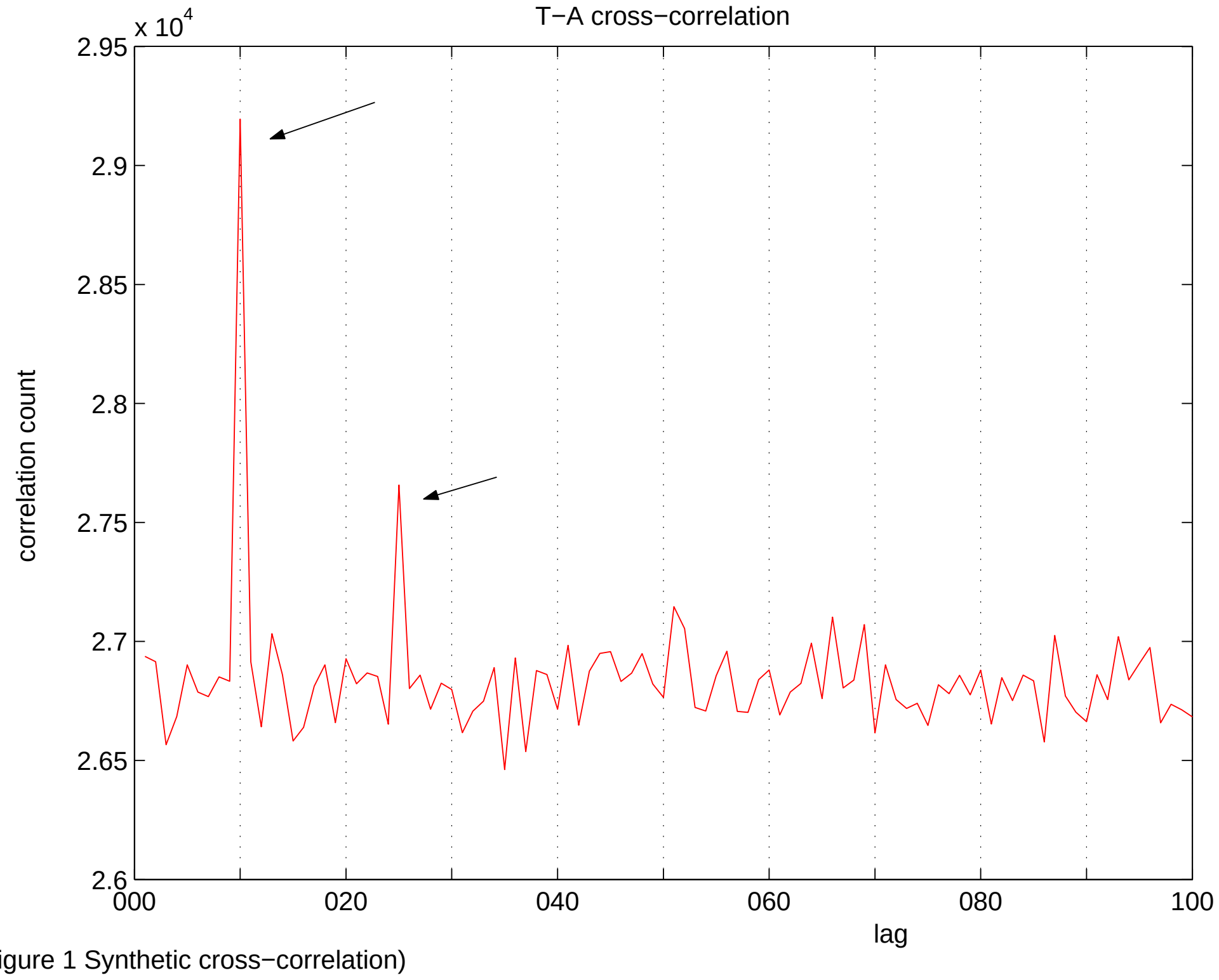

(Figure 1 Synthetic cross−correlation)

##

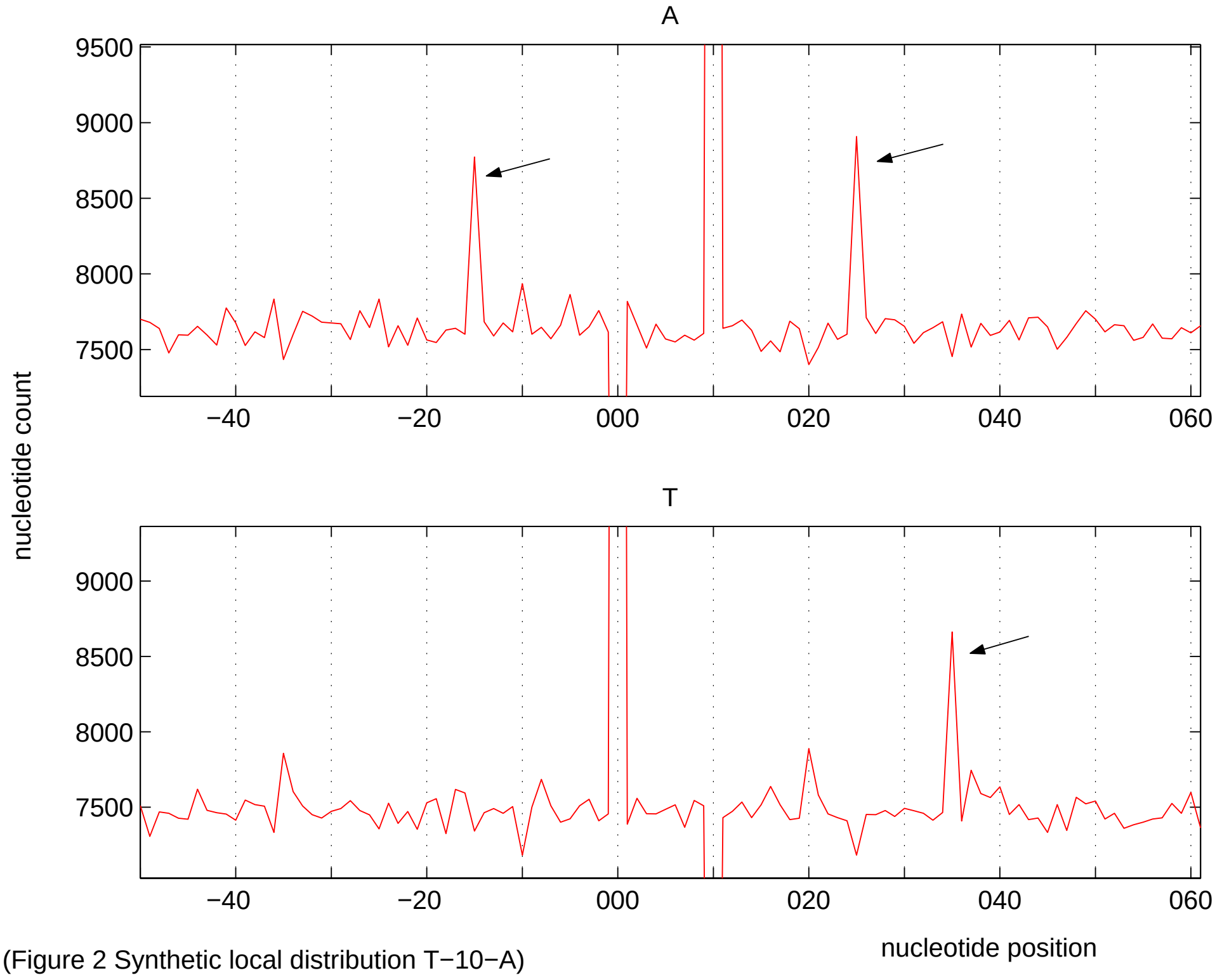

(Figure 2 Synthetic local distribution T−10−A)

##

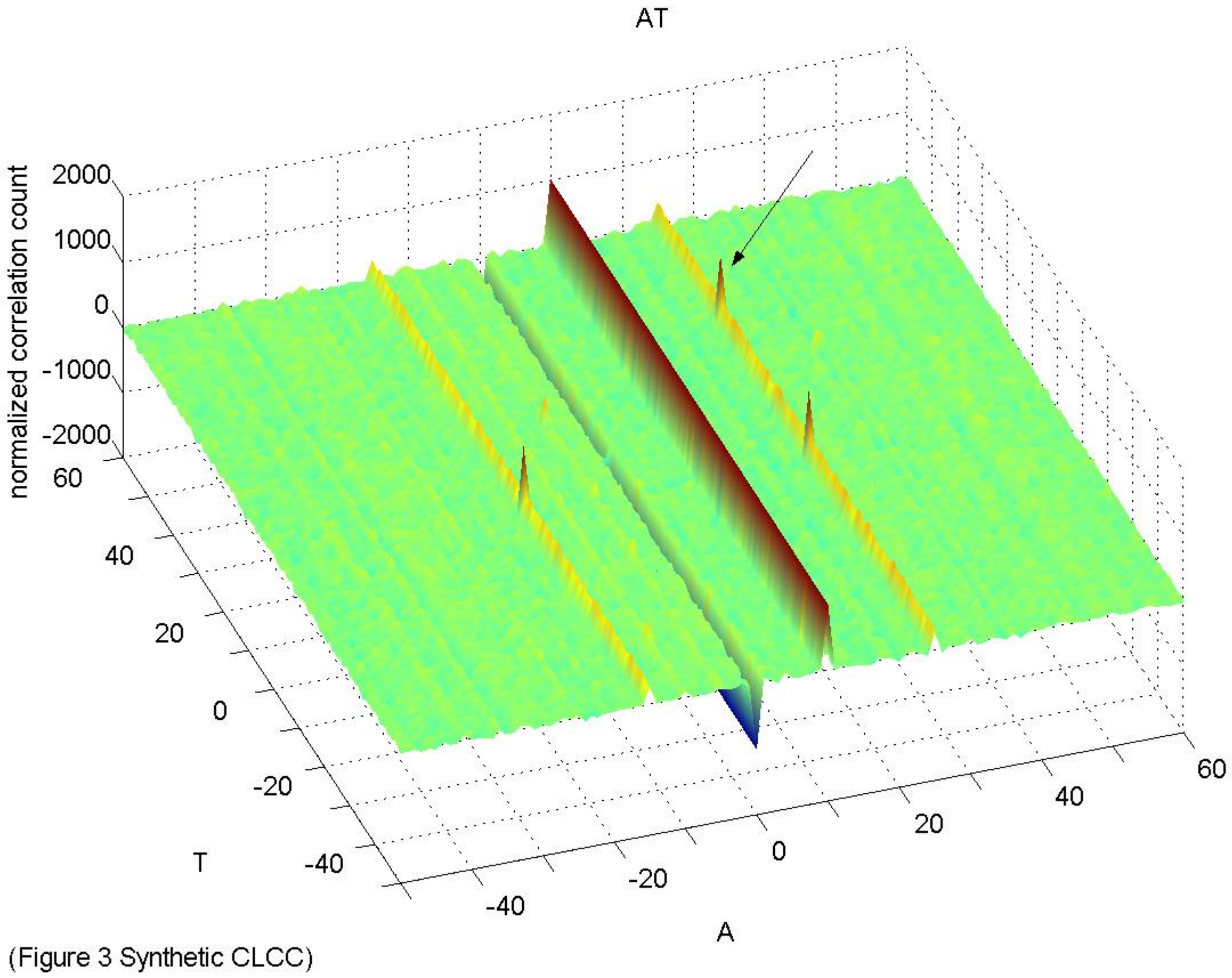

(Figure 3 Synthetic CLCC)

##

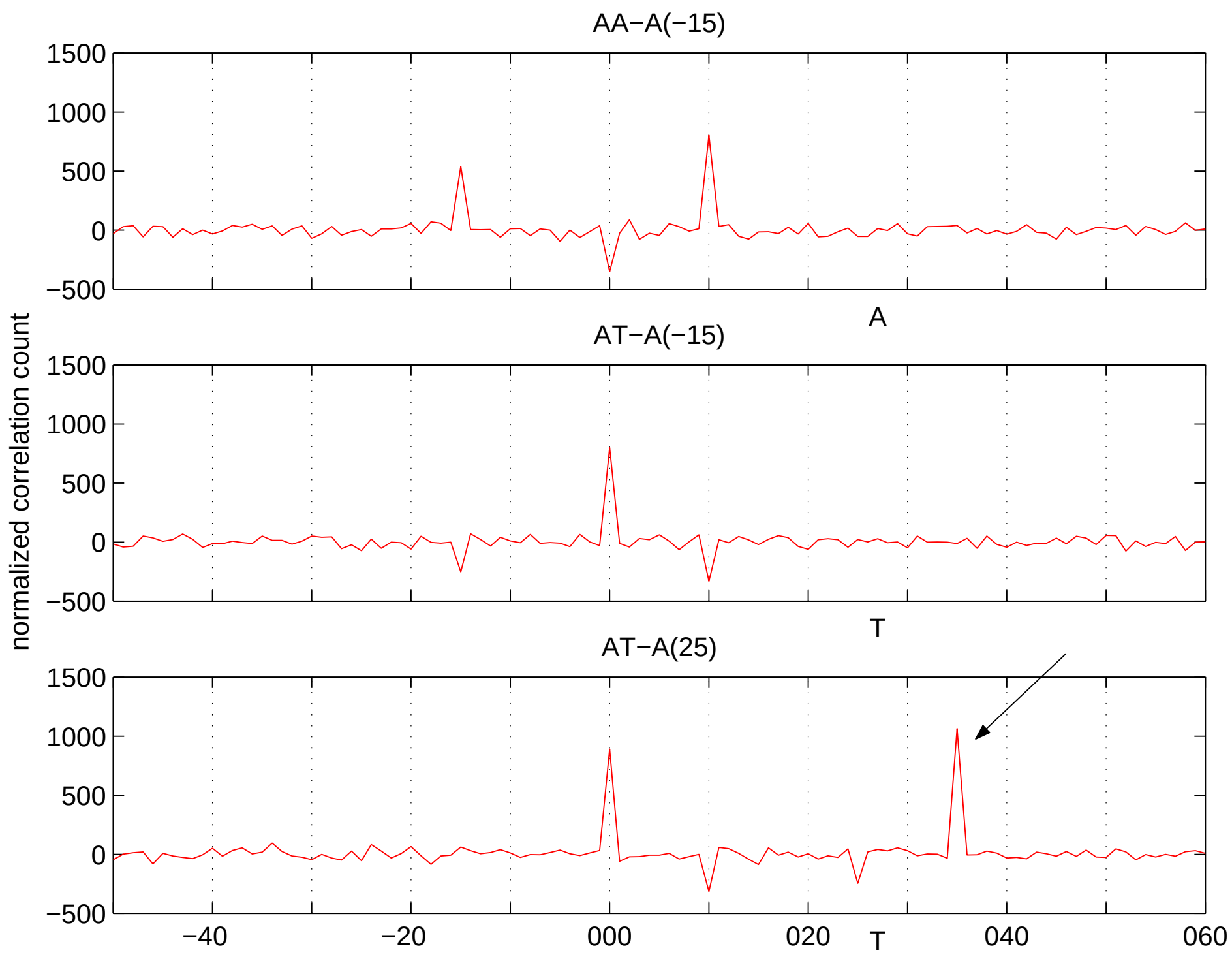

(Figure 4 Synthetic CLCC columns)

##

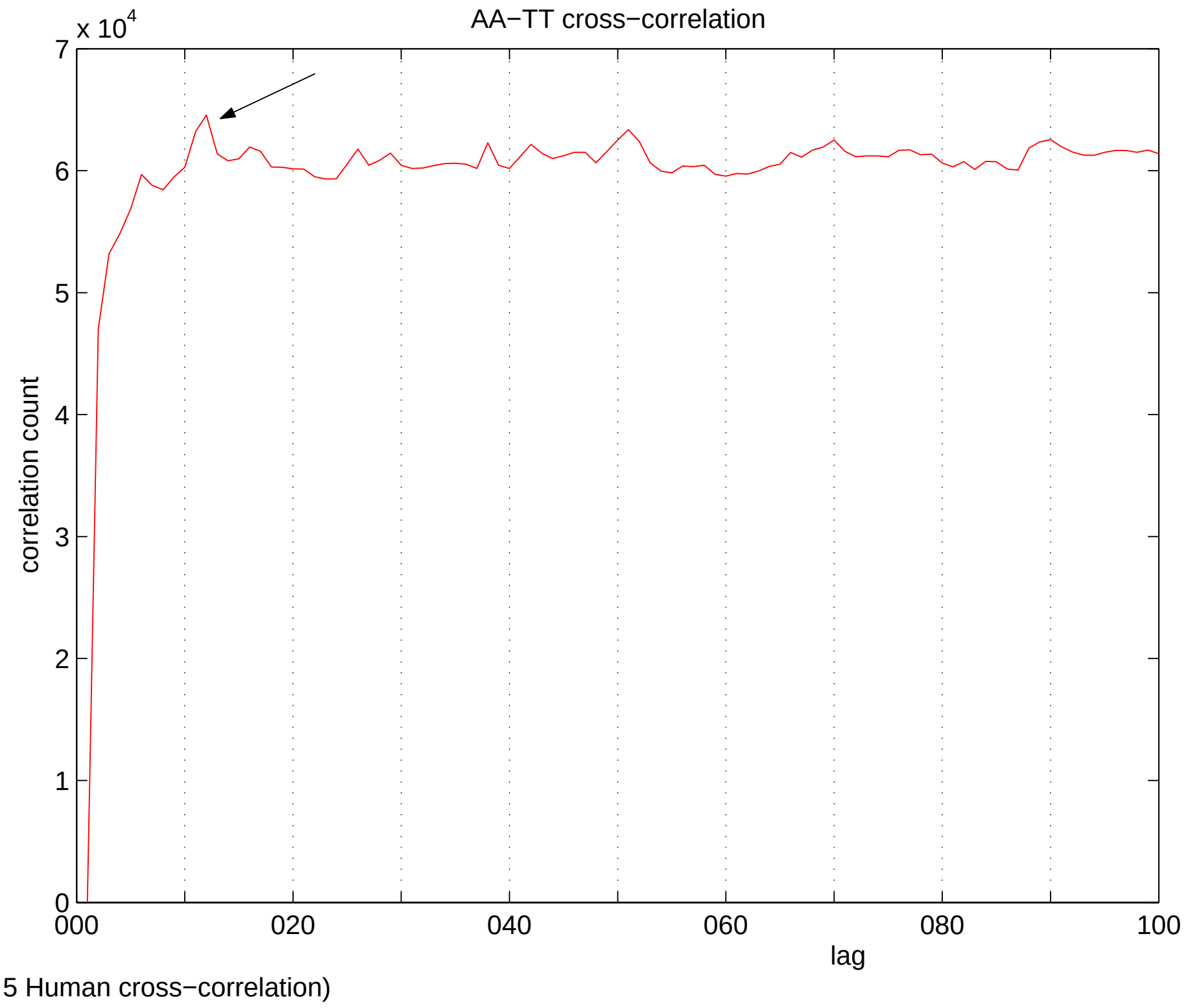

(Figure 5 Human cross−correlation)

##

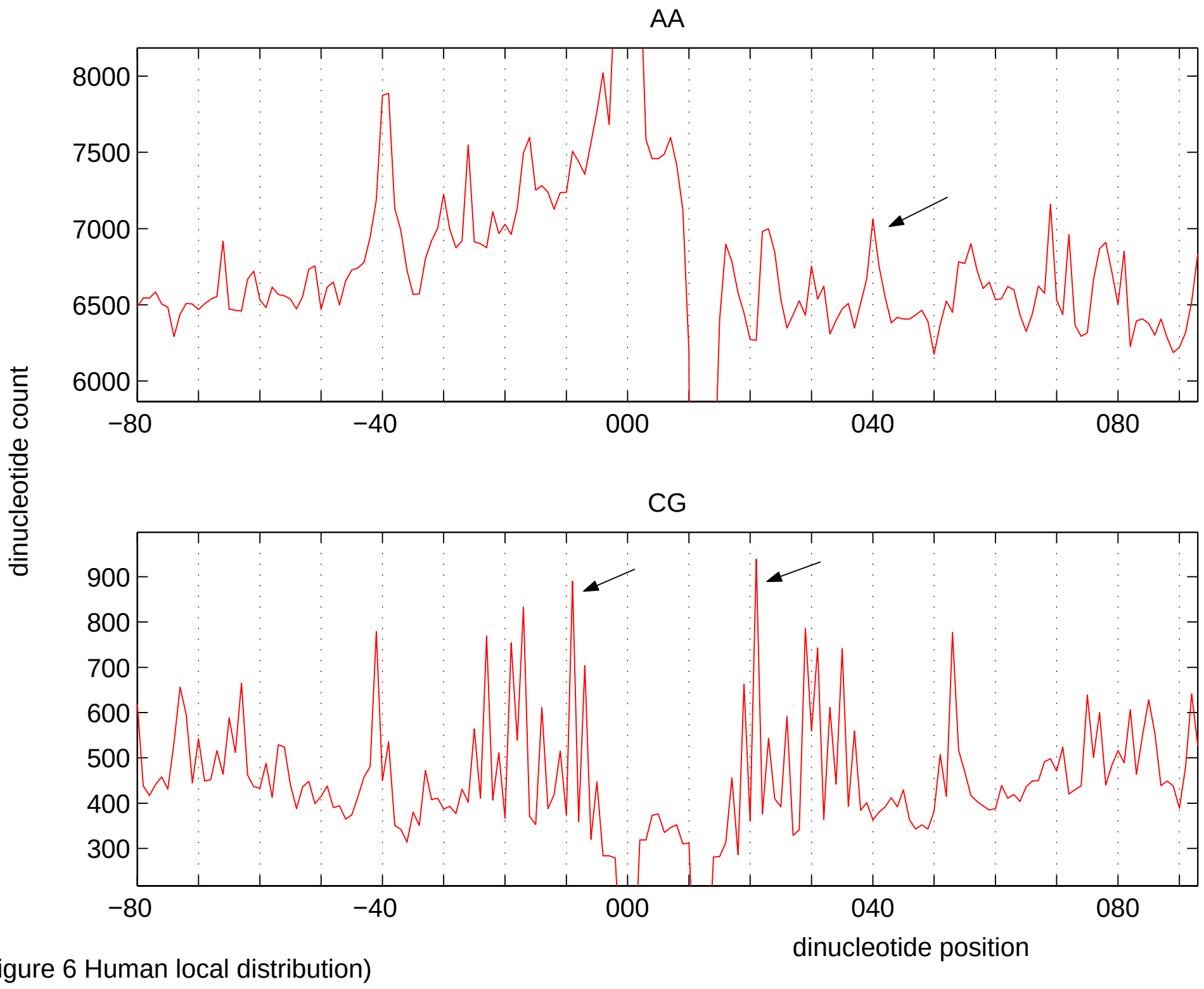

(Figure 6 Human local distribution)

##

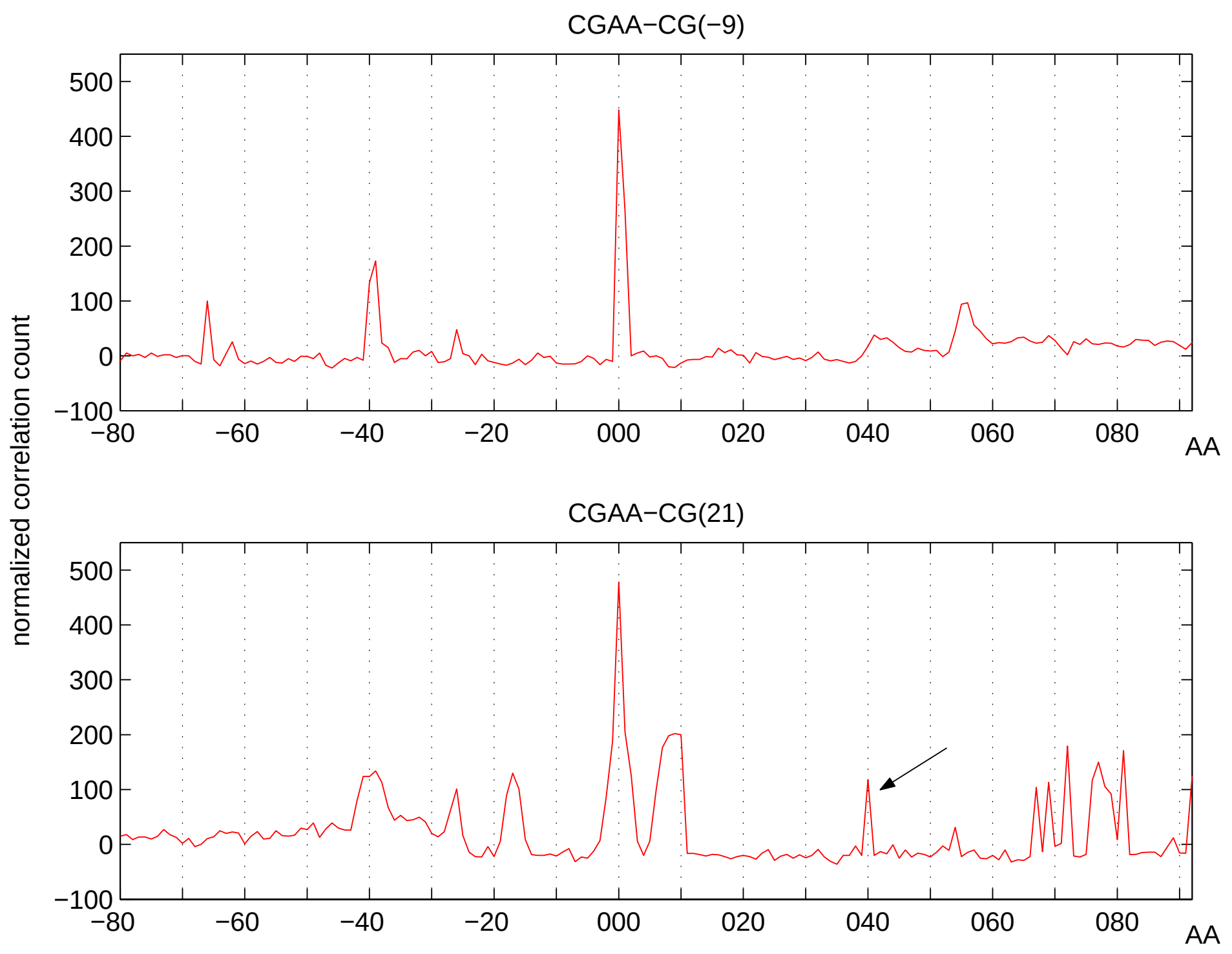

(Figure 7 Human CLCC columns)

##

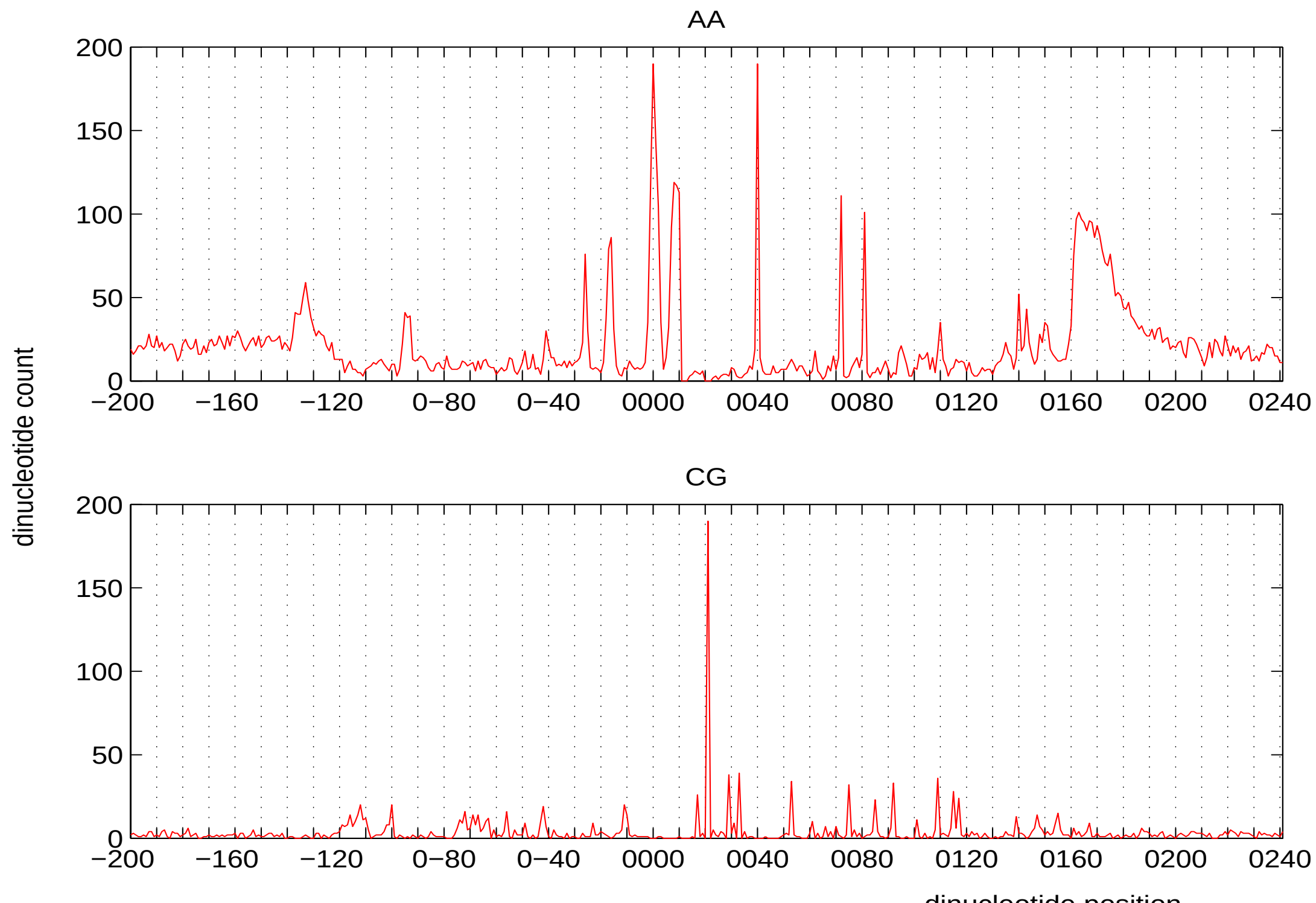

(Figure 8 Human local non−overlapped distribution)

##

Table 1. Alignment with Alu consensus

```
#=======================================================================
# Aligned_sequences: 2
# 1: human-chr1-contig-32--patt-AA-12-TT-9-CG-19-AA-consenSeq
# 2: Alu-Consensus
#=======================================================================

human-chr1-co      1 aaaxxxxxxxxxxggxxxggggxxgggggxcxcxcccxtxaxccccxxxc     50
                                  ||   ||.|  |.||. | | ||. | | |||.   |
Alu-Consensus      1              ggccgggcgcggtggctcacgcctgtaatcccagcac     37

human-chr1-co     51 cttgggggggxxggggggggggxxccxxggggxcxggagttxxagxccx    100
                     .|||||.||..  ||.|||.||.  .|  |.|| | ||||||  || ||
Alu-Consensus     38 tttgggaggccgaggcgggcggatcacctgaggtcaggagttcgagacca     87

human-chr1-co    101 gccxggccaacxxggxgaaaccccxxcccxxcxaaaaatacaaaaattag    150
                     ||| |||||||  || ||||||||  |.|  | |||||||||||||||||
Alu-Consensus     88 gcctggccaacatggtgaaaccccgtctctactaaaaatacaaaaattag    137

human-chr1-co    151 ccgggcgtggtggcgcgcgcctgtaatcccagctactcgggaggctgagg    200
                     ||||||||||||||||||||||||||||||||||||||||||||||||||
Alu-Consensus    138 ccgggcgtggtggcgcgcgcctgtaatcccagctactcgggaggctgagg    187

human-chr1-co    201 caggagaatcgcttgaacccgggaggcggaggttgcagtgagccgagatc    250
                     ||||||||||||||||||||||||||||||||||||||||||||||||||
Alu-Consensus    188 caggagaatcgcttgaacccgggaggcggaggttgcagtgagccgagatc    237

human-chr1-co    251 gcgccactgcactccagcctgggcxacagagxxxgacxcxxxcxcaaaaa    300
                     |||||||||||||||||||||||| ||||||   ||| |   | ||||||
Alu-Consensus    238 gcgccactgcactccagcctgggcgacagagcgagactccgtctcaaaaa    287

human-chr1-co    301 aaaaaaaaaaaaaxa    315
                     |||
Alu-Consensus    288 aaa                290
#=======================================================================
```